\documentclass{PoS}

\title{Rapid variations of polarization in low-mass X-ray binaries}

\ShortTitle{Rapid variations of polarization in LMXBs}

\author{\speaker{David M. Russell}\\
       Astronomical Institute `A. Pannekoek', University of Amsterdam, Science Park 904, 1098 XH, Amsterdam, the Netherlands ~~~~~~~~~~~~~~~~~~~~~~~~~~~~~~~~~~~~~~~~~~       E-mail: \email{D.M.Russell@uva.nl}}

\author{Piergiorgio Casella\footnote{Stand-in speaker.}, ~ Rob Fender\\
        School of Physics and Astronomy, University of Southampton, Highfield, Southampton, England
	}

\author{Paolo Soleri\\
        Kapteyn Astronomical Institute, University of Groningen, PO Box 800, 9700 AV, Groningen, the Netherlands
	}

\author{Magaretha L. Pretorius\\
        European Southern Observatory, Alonso de Cordova 3107, Santiago, Chile
	}

\author{Fraser Lewis\\
        Faulkes Telescope Project, School of Physics and Astronomy, Cardiff University, 5, The Parade, Cardiff, CF24 3AA, Wales; Department of Physics and Astronomy, The Open University, Walton Hall, Milton Keynes, MK7 6AA, England; Division of Earth, Space and Environment, University of Glamorgan, Pontypridd, CF37 1DL, Wales
	}

\author{Michiel van der Klis\\
       Astronomical Institute `A. Pannekoek', University of Amsterdam, Science Park 904, 1098 XH, Amsterdam, the Netherlands\\
	}

\abstract{
Time-resolved optical and infrared polarimetric observations of black hole and neutron star low-mass X-ray binaries are presented. Data were acquired with the VLT, UKIRT and HIPPO on the SAAO 1.9-m. We find that for some sources in outburst, a rapidly variable component of polarization is evident that is stronger in the redder wavebands. We attribute this to the polarimetric signature of synchrotron emission from jets in these systems, the emission of which is known to dominate these redder bands. Such synchrotron emission from jets launched close to black holes and neutron stars can be highly linearly polarized, depending on the configuration of the magnetic field. The variability of the polarization is suggestive of a tangled and turbulent magnetic field at the location of the compact jet. For some sources the position angle of polarization is consistent with a magnetic field that is parallel to the observed radio jet. These are some of the first observational constraints of the geometry and magnetic structure at the inner regions of the outflow. We also present the first ever simultaneous optical polarization and X-ray campaign of an X-ray binary, using data taken simultaneously with HIPPO and RXTE with sub-second time resolution.
}

\FullConference{High Time Resolution Astrophysics (HTRA) IV - The Era of Extremely Large Telescopes\\
                May 5 - 7, 2010 \\
                Agios Nikolaos, Crete Greece}

\begin{document}

\section{Introduction}

A further dimension to high time resolution astrophysics (HTRA), largely unexplored so far \cite{slowet09,pottet10}, is rapid changes in the optical polarization of astronomical sources. However, with new \cite{kanbet03,pottet08,collet09} and future instrumentation, this new region of parameter space may provide vital clues to aid us in understanding the physics of some variable objects. Here, we focus on variability of polarization in low-mass X-ray binaries (LMXBs), and show that rapid variability is present and can be detected in the brightest sources with current instrumentation on moderate size telescopes. A global picture of polarization and magnetic properties of the accretion flows in these objects will come essentially via studies of populations of LMXBs using future facilities that will provide improvements in both sensitivity and temporal domain.

Optical and infrared light from LMXBs can appear polarized either due to intrinsic polarization of the emitting photons, or by the (Thompson or possibly Rayleigh) scattering or absorption of unpolarized photons. Polarization of optical light from LMXBs due to absorption by interstellar dust is inferred by a characteristic \cite{serk73}, unchanging, wavelength-dependent polarization level \cite{schuet04}. Scattering of intrinsically unpolarized thermal emission by electrons \cite{dola84} within the system has been detected as optical polarization that changes as a function of the orbital phase of the binary in two black hole (BH) LMXBs, A0620--00 and GRO J1655--40 \cite{dolata89,glioet98}. The reported strength of the polarization from scattering is typically a few per cent. All other reported measurements of optical polarization in LMXBs have an interstellar dust origin as far as we are aware, except possibly for the BH XTE J1118+480 \cite{schuet04} and the neutron star (NS) LMXB Aql X--1 \cite{charet80}.

Very recently, evidence for polarization intrinsic to the emitting photons of LMXBs has been revealed from infrared observations \cite{shahet08,russfe08}. Shahbaz et al. (2008) \cite{shahet08} performed infrared spectropolarimetry of three LMXBs and found two of them (the NSs Sco X--1 and Cyg X--2) to be intrinsically polarized, with an increasing linear polarization (LP) at lower frequencies (up to $\sim 2$\% in Sco X--1 and $\sim 10$\% in Cyg X--2). This dependence on wavelength cannot be explained by scattering or interstellar dust absorption.

We performed near-infrared (NIR) imaging polarimetry of six LMXBs in 2005--6 using the Infrared Spectrometer And Array Camera (ISAAC) on the Very Large Telescope (VLT) (European Southern Observatory [ESO] programme 275.D-5062) and the UKIRT 1--5 micron Imager Spectrometer (UIST) + Infra-Red Polarimetry facility (IRPOL2) on United Kingdom Infrared Telescope (UKIRT) (making use of the UKIRT Service Observing Programme; UKIRTSERV). The observations and results are summarized in Table 1; this work was published in Russell \& Fender (2008) \cite{russfe08}. Most sources were in quiescence, and no polarization was detected. LP was detected (at the $> 3 \sigma$ level) in two active sources, the NS Sco X--1 and the BH GRO J1655--40 (Fig. 1). The LP is strongest in the reddest NIR filters, and is variable for Sco X--1. For this latter source, data were acquired on four dates and the LP was seen to vary both between dates and within the same night (the time resolution was 2.5 minutes).

\begin{table*}
\caption{Details of the 2005--6 NIR polarimetric observations of LMXBs taken with the VLT + ISAAC (GRO J1655--40) and the UKIRT + UIST (the other five), and main results. 1$\sigma$ errors are given for LP, or 3$\sigma$ upper limits.}
\begin{tabular}{llllll}
\hline
Source&Epoch (days)&Source activity&Filters&LP (filter)\\
\hline

GRO J0422+32 &2006 Aug (1)&Quiescence&J,K&$< 12$\% (J)\\

4U 0614+09   &2006 Oct (1)&Persistent&K&$< 16$\% (K)\\

XTE J1118+480&2006 Feb (2)&Quiescence&J,K&$< 6$\% (J), $< 7$\% (K)\\

Sco X--1    &2006 Feb--Mar (4)&Persistent&J,H,K&0.29--0.65 (J), 0.07--0.57\\
&&&&(H), 0.13--0.91 (K)\\

GRO J1655--40&2005 Oct (2)&Hard state&J,H,K$_{\rm s}$&4.7$\pm$1.5 (H), 5.8$\pm$1.9 (K$_{\rm s}$)\\

Aql X--1     &2006 Aug--Oct (3)&Quiescence&J,H,K&$< 5$\% (J), $< 17$\% (H),\\
&&&&$< 12$\% (K)\\

\hline
\end{tabular}
\end{table*}

\begin{figure}
\includegraphics[width=15cm,angle=0]{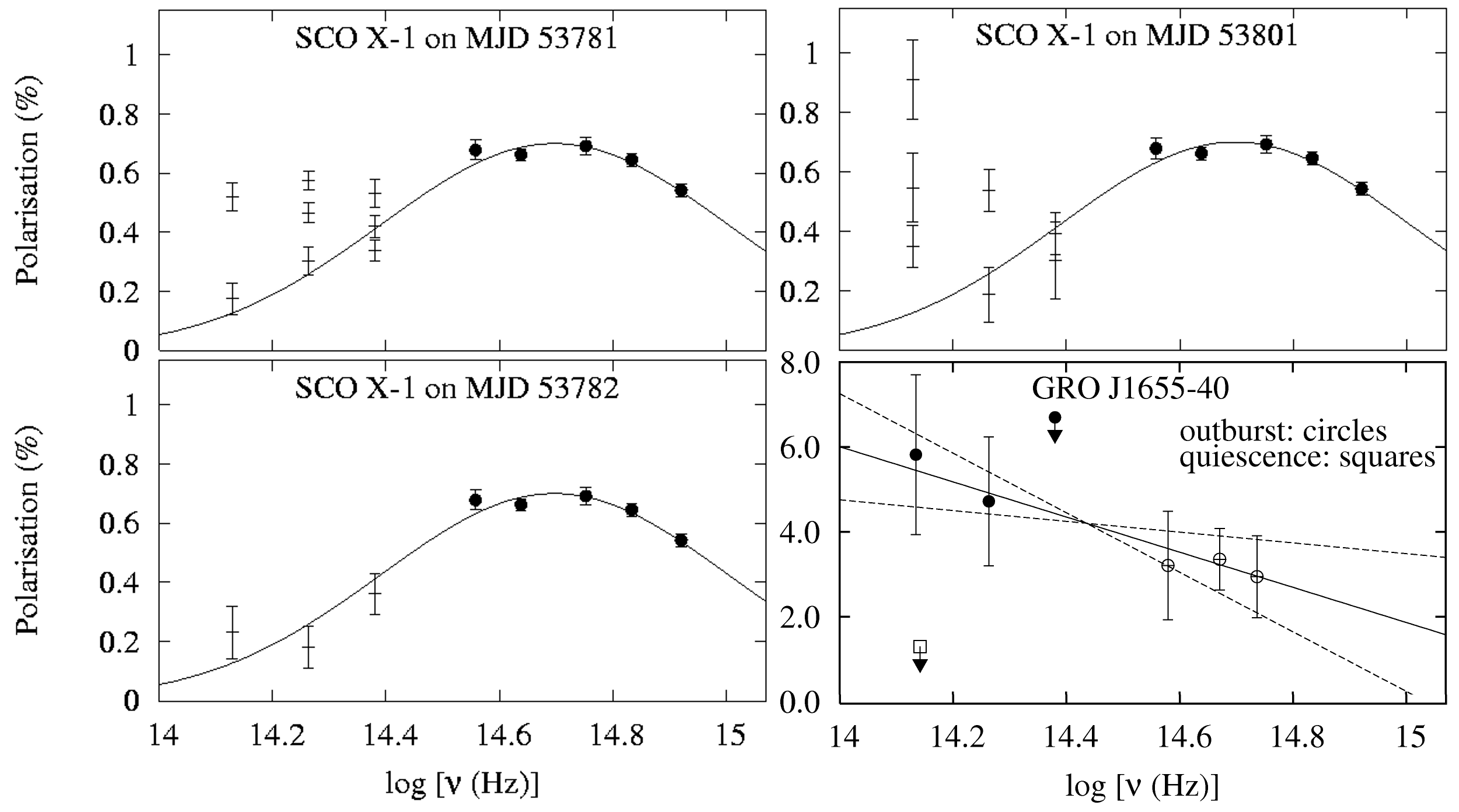}
\caption{NIR intrinsic LP detected in Sco X--1 and GRO J1655--40, as a function of frequency. For Sco X--1 we include the optical data (filled circles) and the fit to these data for an interstellar origin to the polarization (solid curve) from \cite{schuet04}. The crosses on the left with vertical bars are our KHJ-band results (2$\sigma$ detections of LP); a clear variable LP component above that expected from interstellar dust is seen at the lowest frequencies. For GRO J1655--40, VRI-band data are from \cite{glioet98} and a K-band upper limit in quiescence is from \cite{dubuch06}. The lines in the lower right panel show the fit to the outburst data and uncertainty in the slope.}
\end{figure}

\section{The origin of the intrinsic polarization}

Scattering and interstellar absorption cannot reproduce the observed wavelength dependence of the LP in Sco X--1, Cyg X--2 and GRO J1655--40, nor can they explain the short timescale variability of the LP in Sco X--1. Polarized light is however expected from optically thin synchrotron emission from the jets in these systems, which has now been observed in a number of LMXBs in the NIR for both BHs \cite{corbfe02,hyneet03,buxtba04,russet10} and NSs \cite{miglet06,russet07,torret08}. The strength of LP is dependent on the ordering of the magnetic field structure near the emitting region, which for NIR optically thin synchrotron emission is close to the jet launch region at the base of the jet. Thermal emission from the disc dominates more at higher frequencies in the optical, whereas the jet is more dominant in the NIR, so the higher levels of LP in the NIR can be explained by a higher jet/disc flux ratio in this region of the spectrum. Moreover, the position angle (PA) of the polarization in Sco X--1 (measured by us and by Shahbaz et al. 2008 \cite{shahet08}) is consistent with the magnetic field being aligned with the axis of the jet, which is known from radio observations \cite{fomaet01}. We also considered LP caused by the scattering of the jets on the companion star surface, which can be ruled out because this would result in LP only at some phases of the binary orbit, which is not observed \cite{russfe08}.

It is for these reasons that both us and Shahbaz et al. interpret this polarization as the first detections of the polarized inner regions of the jets. For a highly ordered magnetic field, the LP from optically thin synchrotron emission could be as high as $\sim 70$\% \cite{rybili79,bjorbl82}. The levels detected (a few per cent at most) suggest a tangled magnetic field is present, unless the majority of the NIR emission is produced in the (unpolarized) disc or star. The disc and/or star do seem to contribute or even dominate the NIR for the three sources \cite{miglet07,russet07}, so $f$, the dimensionless parameter representing the ordering of the magnetic field (LP $= 0.7 \times f$), cannot be constrained for Sco X--1 and Cyg X--2 because the jet/disc flux ratio is unknown. For GRO J1655--40 we estimate the jet/disc ratio from broadband spectral energy distributions \cite{miglet07} and infer $f = 0.41\pm 0.19$ \cite{russfe08}. To constrain the value of $f$ further, we require data during times in which the jet is known to be dominating the emission.

\begin{figure}
\includegraphics[width=7.5cm,angle=0]{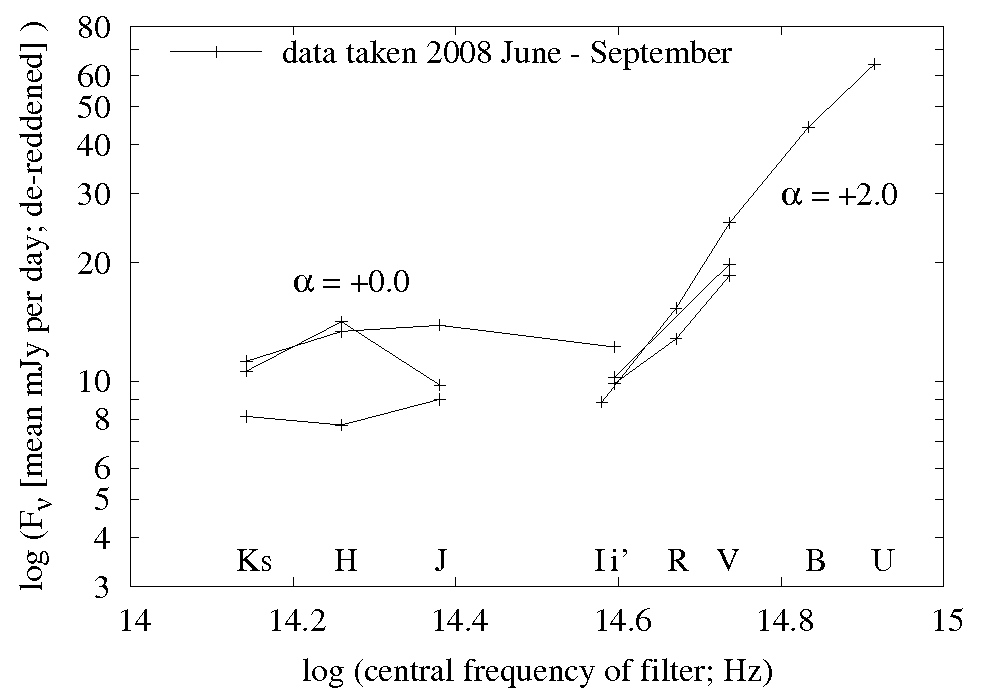}
\includegraphics[width=7.65cm,angle=0]{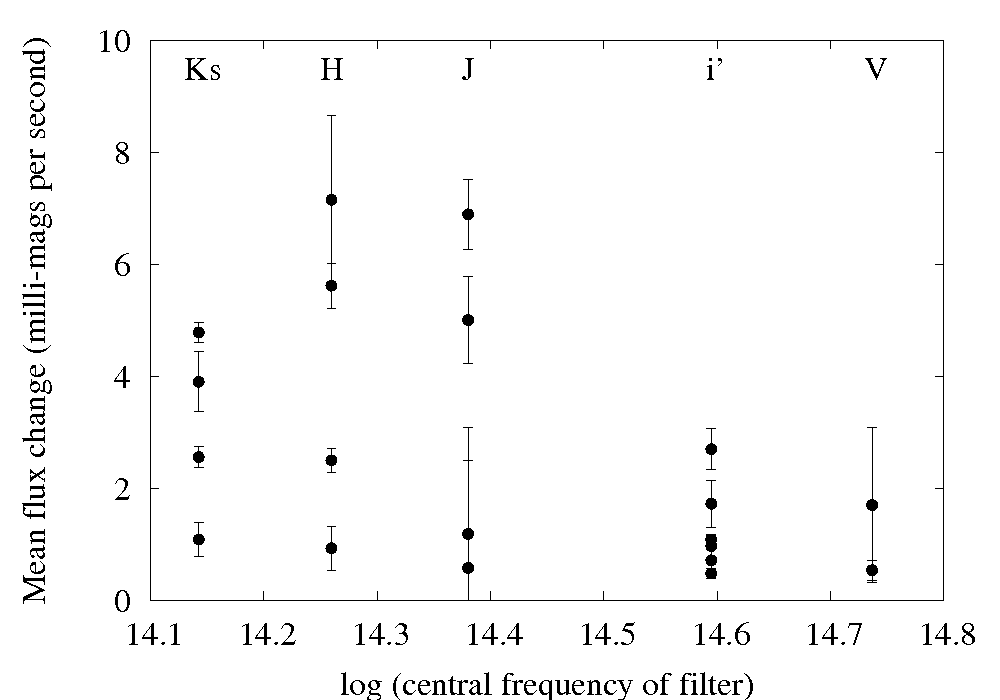}
\caption{\emph{Left panel:} Optical--NIR spectral energy distributions of GX 339--4 during a low luminosity hard state in 2008 June -- September. Data were acquired with ISAAC on the VLT, the Faulkes Telescope South and the SAAO 1.9m Telescope. The data were de-reddened assuming an extinction of A$_{\rm V}=3.9$ mag \cite{jonket04} and following the extinction law of \cite{cardet89}. \emph{Right panel:} The rapidity of the variability from the Faulkes and VLT observations (the mean for each epoch in each filter), as a function of frequency of the filter (see text).}
\end{figure}

\section{New NIR polarimetry of GX 339--4}

In 2008 September we observed the BH GX 339--4 on four dates (4$^{th}$, 8$^{th}$, 17$^{th}$ and 25$^{th}$) with ISAAC on the VLT (ESO Director's Discretionary Time programme 281.D-5029) in imaging polarimetry mode. At the time, the NIR and optical emission was highly variable (with high amplitude rms variability) and was correlated with the X-ray flux; the variability was found to originate in the jets \cite{russet08,caseet10}. In Fig. 2 (left panel) we show the optical--NIR spectral energy distributions (SEDs) of GX 339--4 during this highly variable `hard X-ray state' during which a steady compact jet is expected to exist (for state definitions related to jets see e.g. \cite{gallet03,fendet09,bell10}). Data were acquired with ISAAC on the VLT, the Faulkes Telescope South (as part of a long-term monitoring campaign; Lewis et al., in preparation) and the South African Astronomical Observatory (SAAO) 0.75 metre telescope. The data were de-reddened assuming an extinction of A$_{\rm V}=3.9$ mag \cite{jonket04} and following the extinction law of \cite{cardet89}.

Two components of emission are evident; one with a blue spectrum, $\alpha \approx +2.0$ (where $F_{\nu} \propto \nu^{\alpha}$; possibly thermal emission from X-ray reprocessing on the accretion disc surface) and one with an approximately flat spectrum, $\alpha \approx 0.0$ (from synchrotron emission from the jet; see also \cite{corbfe02,homaet05,caseet10}). It is clear from these SEDs that the jet/disc flux ratio increases at lower frequencies. By extrapolating the average SED of the blue component into the NIR, we estimate that it contributes $\sim 24$\%, $\sim 11$\% and $\sim 6$\% of the average J, H and K$_{\rm s}$-band flux densities (however this is sensitive to the value of the extinction A$_{\rm V}$, which is uncertain \cite{buxtve03,jonket04}). By subtracting this contribution from the J, H and K$_{\rm s}$ flux densities we estimate the jet to have mean flux densities of $\sim 8.2$ mJy in J, $\sim 10.5$ mJy in H and $\sim 9.3$ mJy in K$_{\rm s}$. The spectral index of the jet between J and H is $\alpha \approx -0.9$, consistent with optically thin synchrotron, and between H and K$_{\rm s}$ it is $\alpha \approx +0.4$, an inverted spectrum inconsistent with optically thin synchrotron and more consistent with optically thick synchrotron (similar to one SED presented in \cite{corbfe02}).

In addition, the variability is strongest in the NIR J and H-bands (Fig. 2 right panel). The rapidity is defined as the mean of the differences between consecutive magnitudes, divided by the time resolution. The error bars represent the measured variability of a field star, which should be close to zero if noise is minimal. We do not include data in which the field star is measured to vary by $> 1$ milli-magnitudes per second. In the NIR J and H-bands, GX 339--4 varied by up to 7 milli-magnitudes per second (up to 5 in K$_{\rm s}$) whereas the maximum measured from the optical V and i' bands on six separate dates was $\leq 3$ milli-magnitudes per second (a similar trend was found by \cite{gandet10}). We expect the optically thin synchrotron emission to be highly variable and the optically thick emission less so, since the latter is a superposition of overlapping synchrotron components, which will dilute both the variability and polarization. Both the SED of the jet in the NIR and the wavelength-dependent variability are therefore consistent with the break (turnover) in the jet spectrum, between optically thick and thin emission, to reside around the H-band. This result is similar to other works of optical--NIR studies of GX 339--4 \cite{corbfe02,coriet09}.

In Fig. 3 we present the NIR polarimetric results. The Stokes parameters q and u
\newline
 ( $LP = \sqrt{q^2+u^2}$ ) of GX 339--4 and a close field star are derived by correcting the measured values for instrumental polarization using the unpolarized standard star WD 1620--391. The LP is low for both GX 339--4 (LP $\leq$ 2.8\%) and the field star (LP $\leq$ 1.3\%). The values of q and u are clustered around zero approximately uniformly for the field star, which is expected for an $\sim$ unpolarized source with some low level of noise. However for GX 339--4 there is a positive correlation between q and u. Using the Spearman's Rank method we find the correlation is significant at the 3.4$\sigma$ level and it also passes through (0,0); the fit and 1$\sigma$ error is u $= (0.68 \pm 0.08)$ q $+ (0.002 \pm 0.001)$. This corresponds to a PA of polarization ( $tan [2 PA] = \frac{u}{q}$ ) of $(17.0 \pm 1.6)^\circ$.
 
Since we know the jet dominates the NIR emission during these observations and the jet is variable on short timescales, we conclude that the polarization signature of the jet is detected, and changes rapidly. For optically thin synchrotron emission, the PA is parallel to the electric vector and perpendicular to the magnetic field vector. The measured polarization PA differs by $(81 \pm 4)^\circ$ with the PA of the resolved radio jet of $(-64 \pm 2)^\circ$ \cite{gallet04}, so the magnetic field near the base of the jets in GX 339--4 is $\sim$ parallel to the jet axis.

On each date four measurements of LP were made (one in each of the four filters SZ, J, H and K$_{\rm s}$); the time resolution is on the order of 3 -- 4 minutes. If the emission is purely optically thin synchrotron, which it likely is in the H--J region of the SED (see above) then we estimate the ordering of the magnetic field to be $f \leq 0.04$ on these timescales. This is very low, and provides vital observational clues to the structure of the inner accretion (out)flows of BHs, which could be used to compare to those predicted by simulations and models of jet formation (these results will be published in Russell et al., in preparation). It is worth noting that it was predicted in 1982 \cite{fabiet82} that the optical flares in GX 339--4 may be polarized; now almost 30 years later this has been confirmed.

\begin{figure}
\includegraphics[width=13cm,angle=0]{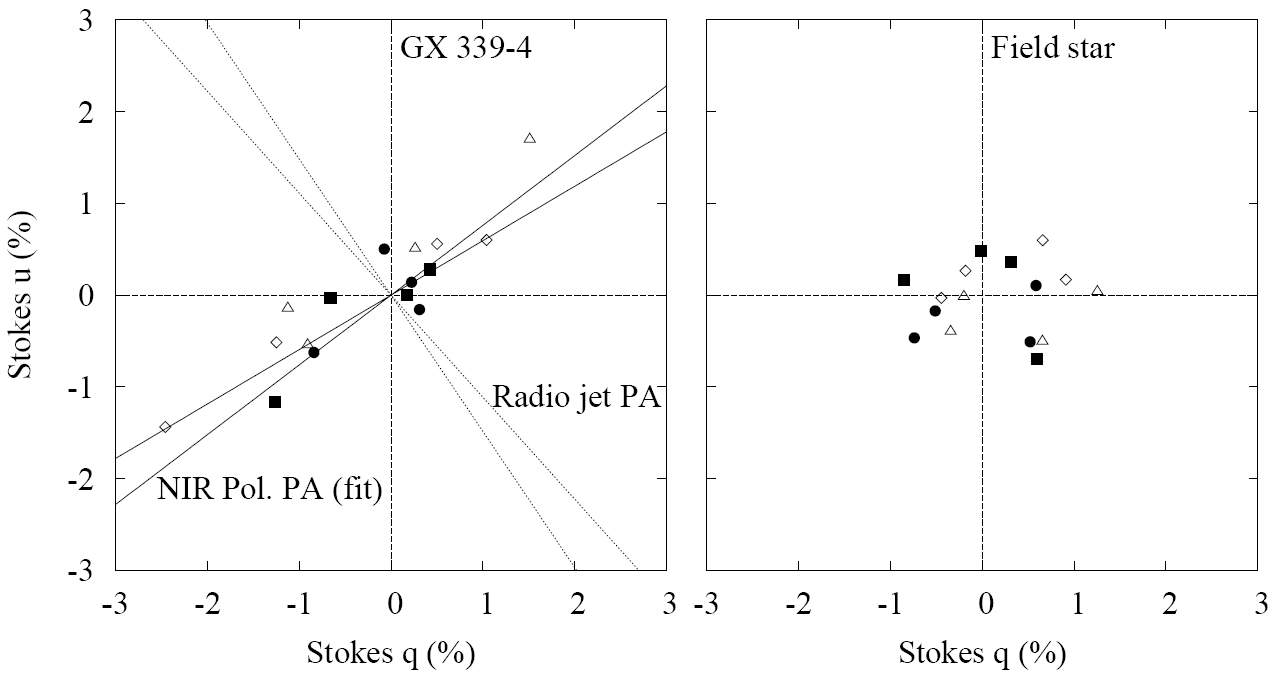}
\caption{\emph{Left panel:} q and u Stokes parameters of GX 339--4, corrected for filter-dependent instrumental polarization (filled circles = K$_{\rm s}$-band, open diamonds = H, filled squares = J, open triangles = SZ). There is a correlation between q and u; the fit to the correlation (1$\sigma$ errors) is shown (`NIR Pol. PA (fit)'), which corresponds to a PA of polarization of $(17.0 \pm 1.6)^\circ$. We also plot the correlation that corresponds to the measured PA of the resolved radio jet of $(-64 \pm 2)^\circ$ \cite{gallet04} (`Radio jet PA'). \emph{Right panel:} q and u Stokes parameters of a close by field star, corrected for filter-dependent instrumental polarization.}
\end{figure}

\section{New optical fast timing polarimetry of Sco X--1 with simultaneous X-ray}

We observed the NS Sco X--1 on two nights in February 2009 with the two channel HIgh speed Photo-POlarimeter (HIPPO) \cite{pottet08,pottet10} mounted on the 1.9-m optical telescope of the South African Astronomical Observatory (SAAO). With HIPPO, Stokes parameters are recorded every 0.1 sec and photometry every 1 millisecond, simultaneously in both channels. We used V and I-band filters for the two channels in order to achieve simultaneous flux and polarimetry at two wavelengths.

\begin{figure}
\includegraphics[width=7.5cm,angle=0]{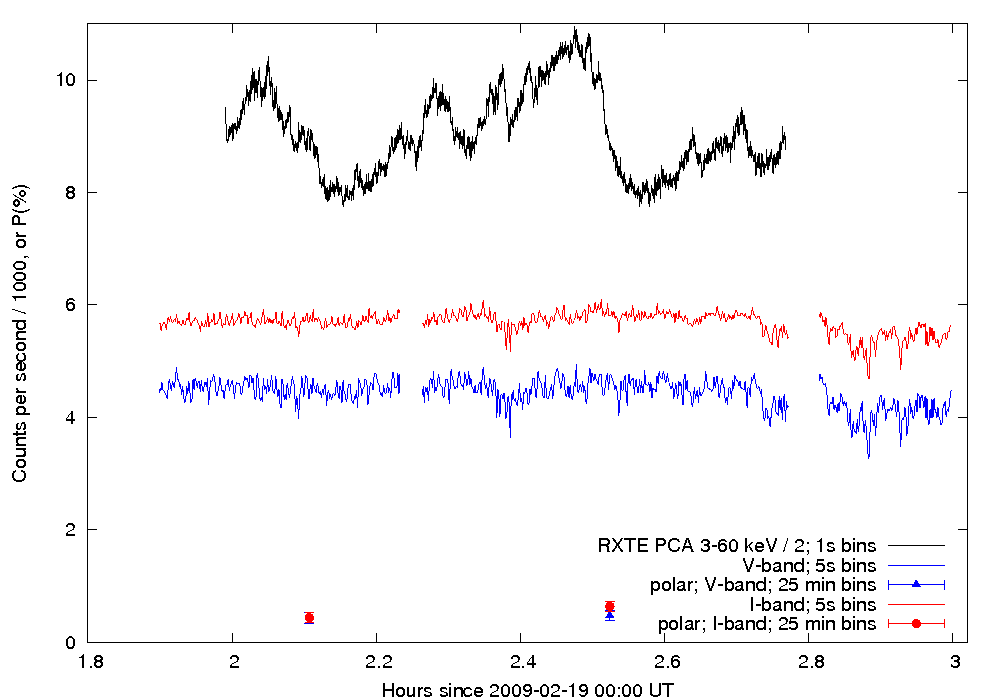}
\includegraphics[width=7.5cm,angle=0]{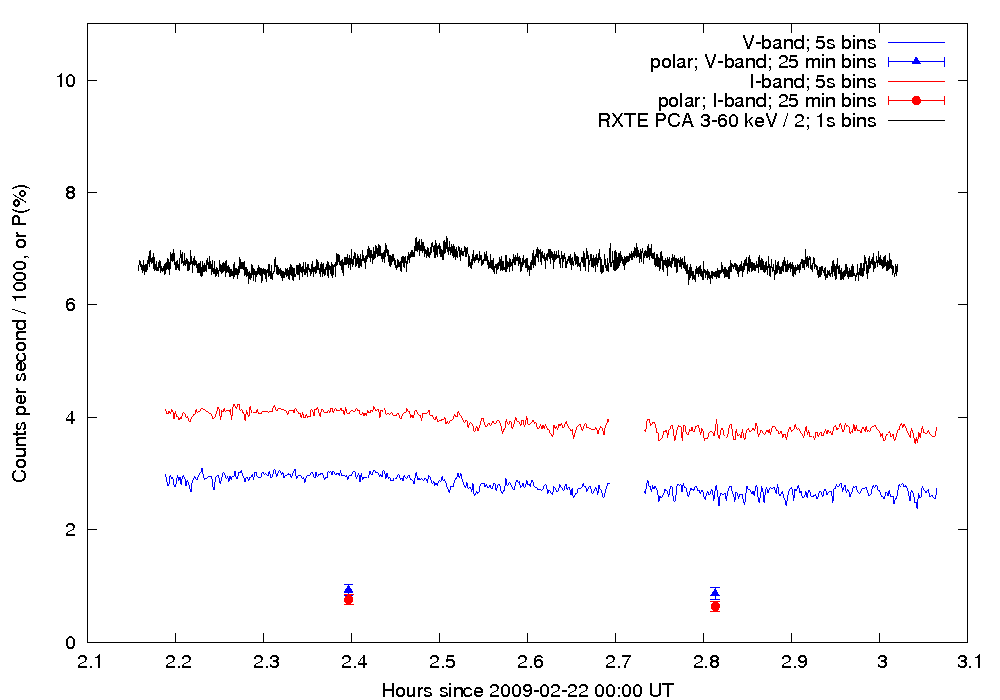}\\
\includegraphics[width=7.5cm,angle=0]{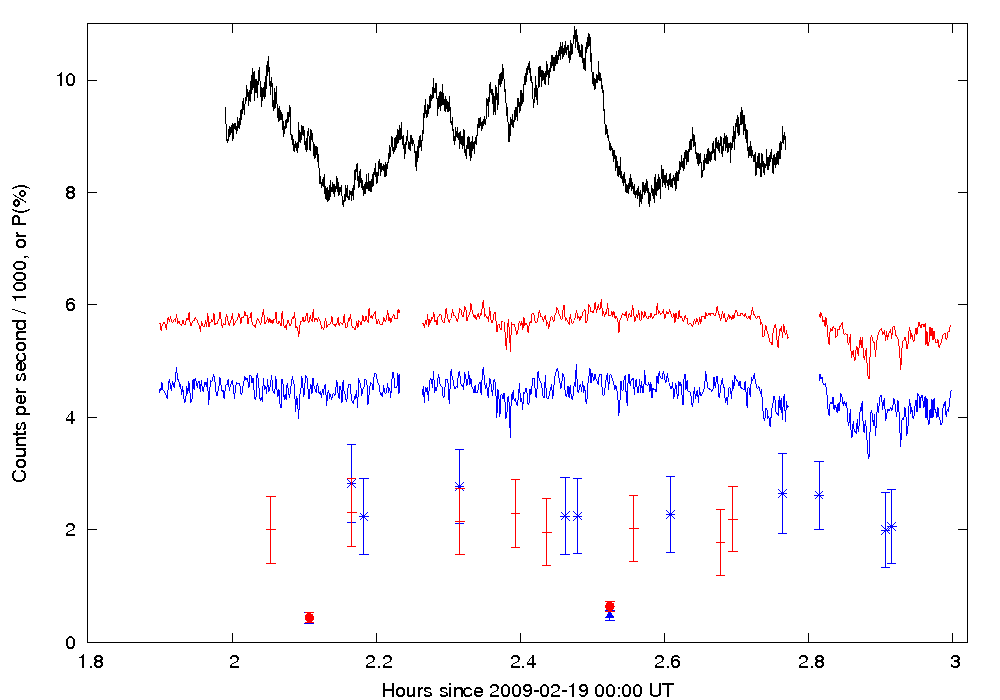}
\includegraphics[width=7.5cm,angle=0]{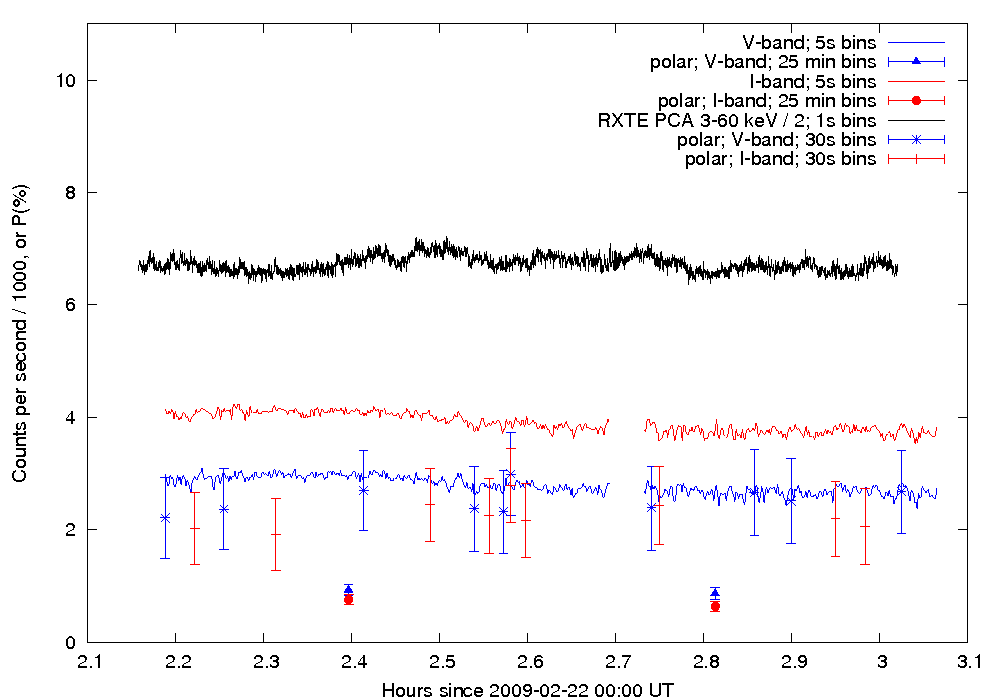}\\
\includegraphics[width=7.5cm,angle=0]{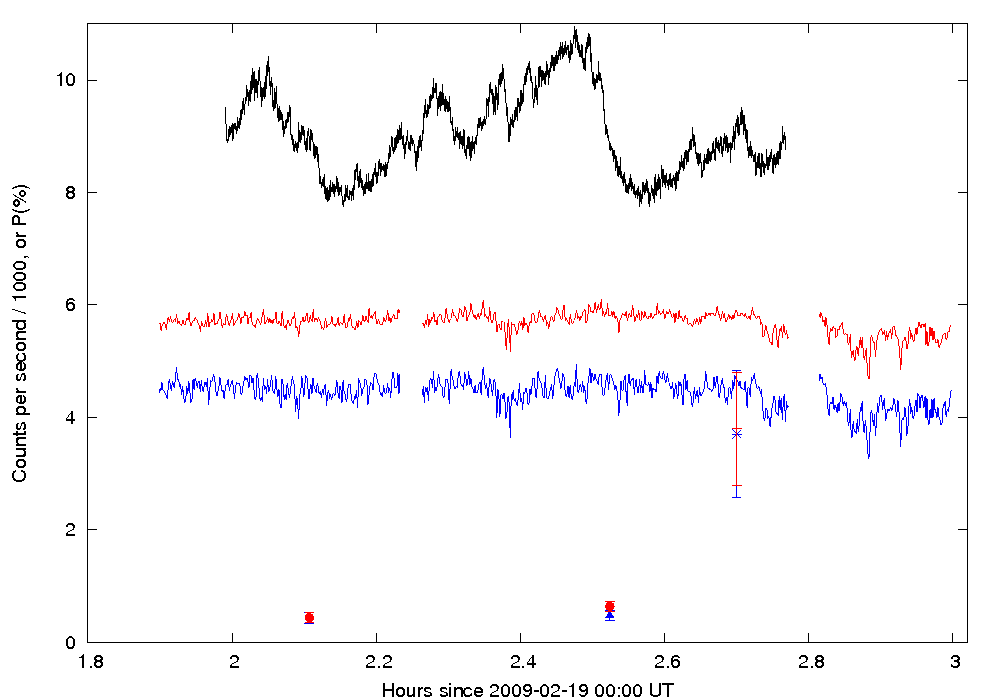}
\includegraphics[width=7.5cm,angle=0]{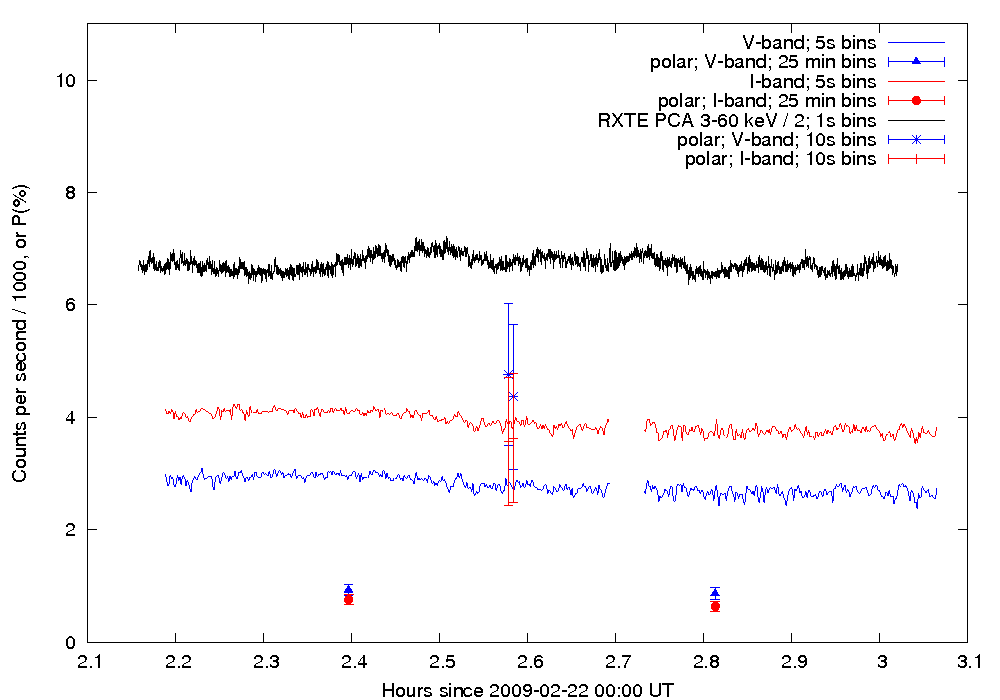}
\caption{Simultaneous light curves: optical flux and polarization in two channels (V-band in blue and I-band in red; from HIPPO) and X-ray 3--60 keV flux per PCU (in black; RXTE PCA). \emph{Left panels:} Data taken on 2009-02-19. The time resolution of the optical flux and X-ray flux is 5 seconds and 1 second, respectively. The polarization measurements are made taking data over timescales of 25 minutes (`25 min bins'; top panel), 30 seconds (`30s bins'; only 3$\sigma$ detections of polarization in one of the two channels are shown; centre panel) and 10 seconds (`10s bins'; only 3$\sigma$ detections of polarization in both channels simultaneously are shown; lower panel). The X-ray light curve and `25 min bins' optical polarization data are also shown in the centre and lower panels for comparison. \emph{Right panels:} Same as the left panels but for data taken on 2009-02-22.}
\end{figure}

X-ray observations of Sco X--1 using the Proportional Counter Array (PCA) aboard the Rossi X-ray Timing Explorer (RXTE) satellite were acquired on seven consecutive dates, 19--25 Feb. On 19$^{th}$ and 22$^{nd}$ Feb, simultaneous optical--X-ray observations were arranged. This is the first time as far as we are aware, that optical polarimetry and X-ray observations of a LMXB have been performed simultaneously. Z-sources display a number of X-ray states which form a characteristic pattern in an X-ray colour--colour diagram or hardness--intensity diagram \cite{hasiva89}. The states (the `horizontal branch', the `normal branch' and the `flaring branch') occupy different regions of these diagrams. It has been found that the behaviour of the radio jets correlates with the X-ray state. According to the unification of radio jets in NS Z-sources \cite{hjelet90,miglfe06}, the jets are absent in the flaring branch, present in the horizontal branch and discrete ejections are launched in the normal branch.

The X-ray and optical flux (and polarimetric) light curves 
are plotted in Fig. 4. The light curves from 19$^{th}$ and 22$^{nd}$ Feb are in the left and right panels, respectively. Comparing the two upper panels, the X-ray and optical fluxes were both brighter and more variable on 19$^{th}$ compared to 22$^{nd}$. Integrating over 25 minutes, the LP is measured to be $< 1$ \% (also shown in the upper panels) but significant, and consistent with an interstellar origin (see Section 2 and Fig. 1). In the centre two panels of Fig. 4 we plot measurements of polarization but binned over only 30 seconds. With smaller bins the error bars are larger, but we plot just the 3$\sigma$ detections of polarization in these centre panels. We expect some false 3$\sigma$ detections; below we discuss whether (some of) these detections are real or are due to noise. LP is again plotted in the lower panels, this time from just 10-second bins and we only plot the data that are 3$\sigma$ detections \emph{in both channels simultaneously}. In the lower right panel (22$^{nd}$ Feb), there appears to be a polarization flare lasting 2 bins ($\sim$ 20 sec).

\begin{figure}
\includegraphics[width=6.1cm,angle=0]{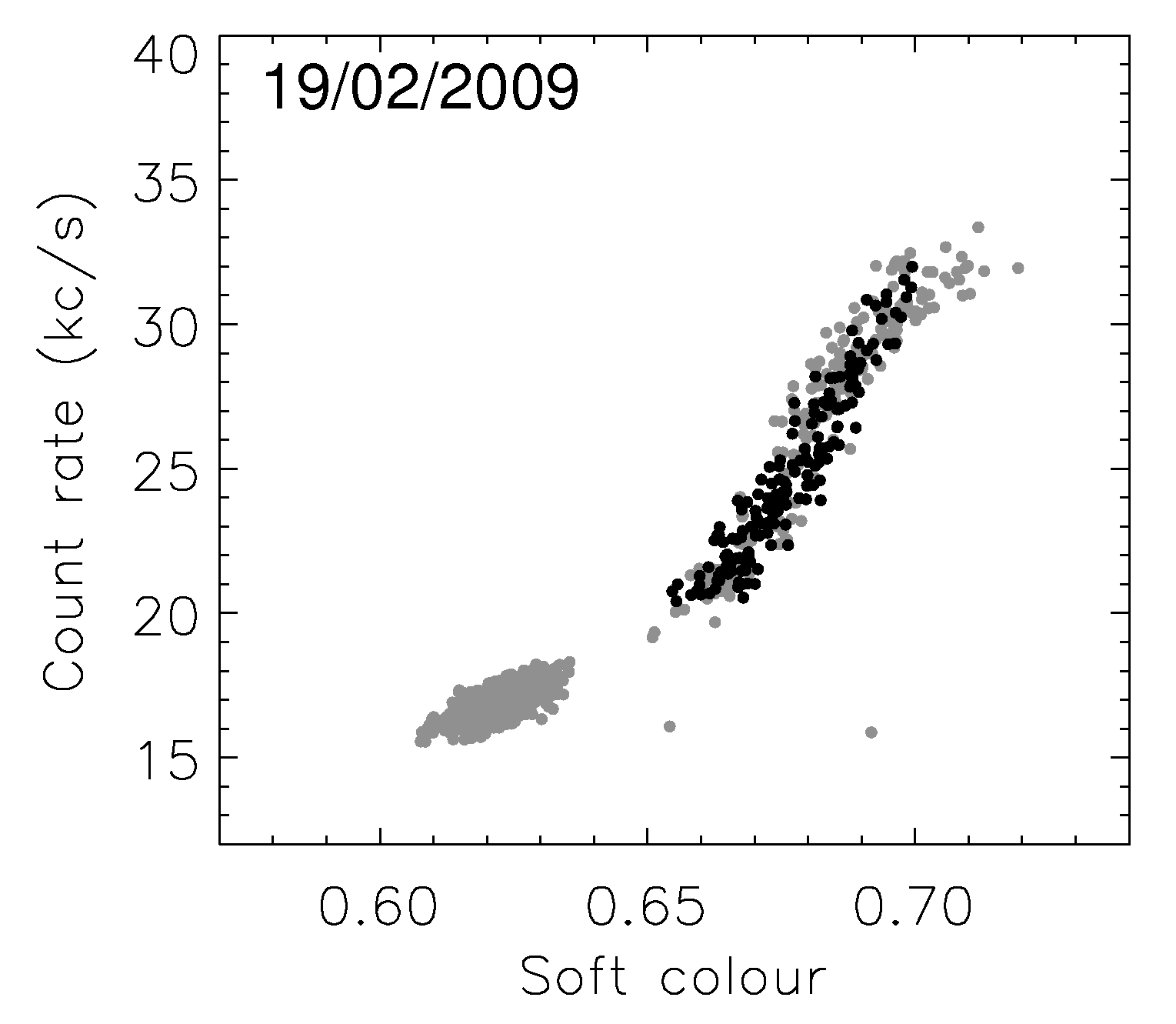}
\includegraphics[width=6cm,angle=0]{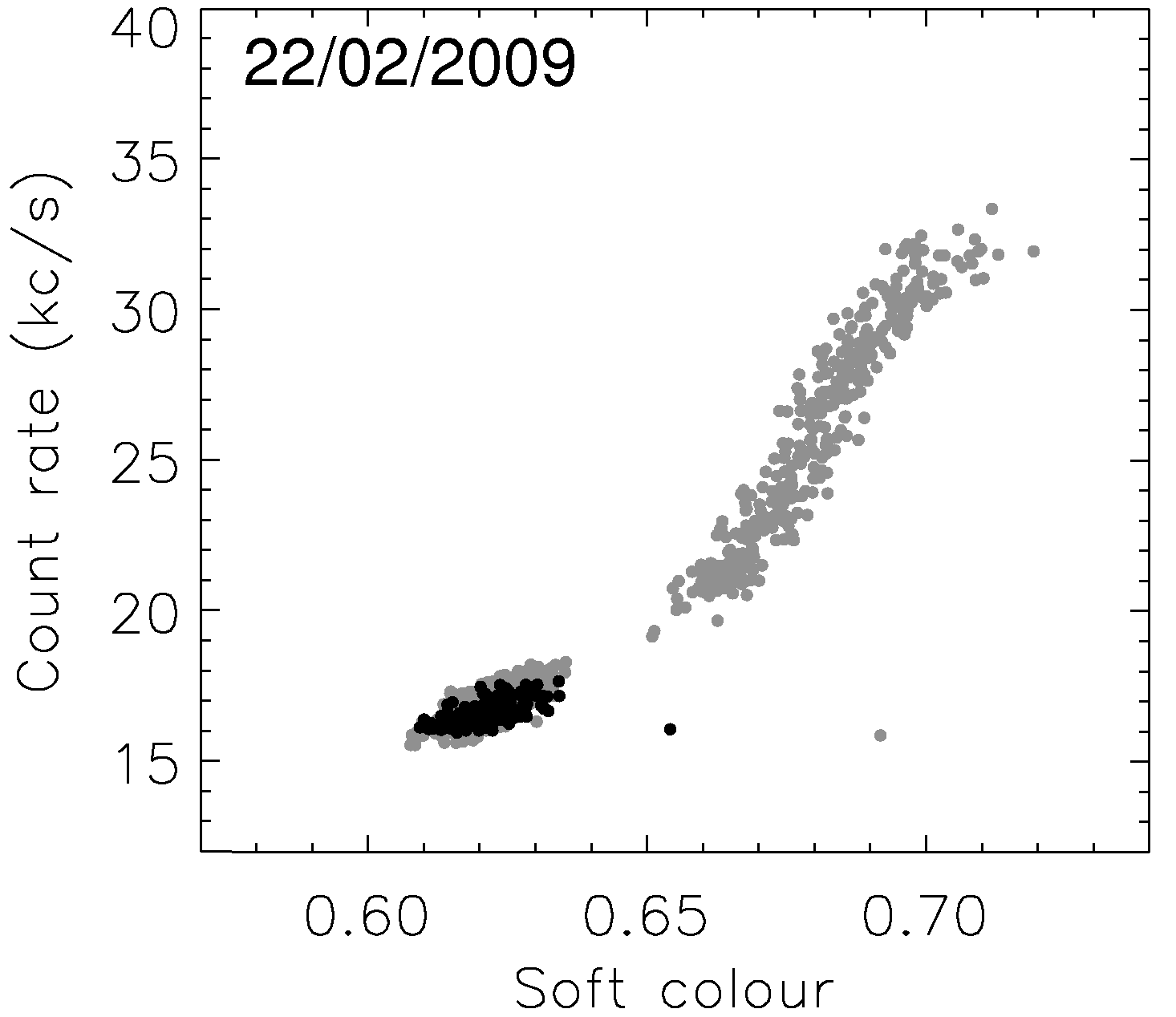}
\caption{X-ray hardness--intensity diagram for Sco X--1. The count rate is in the 3--60 keV range and the soft colour is defined by the 4.4--6.2 keV flux divided by the 3.0--4.4 keV flux. All data from the seven dates are shown in grey, and the data on 19$^{th}$ and 22$^{nd}$ Feb are shown in black in the left and right panels, respectively.}
\end{figure}

\begin{figure}
\includegraphics[width=7.5cm,angle=0]{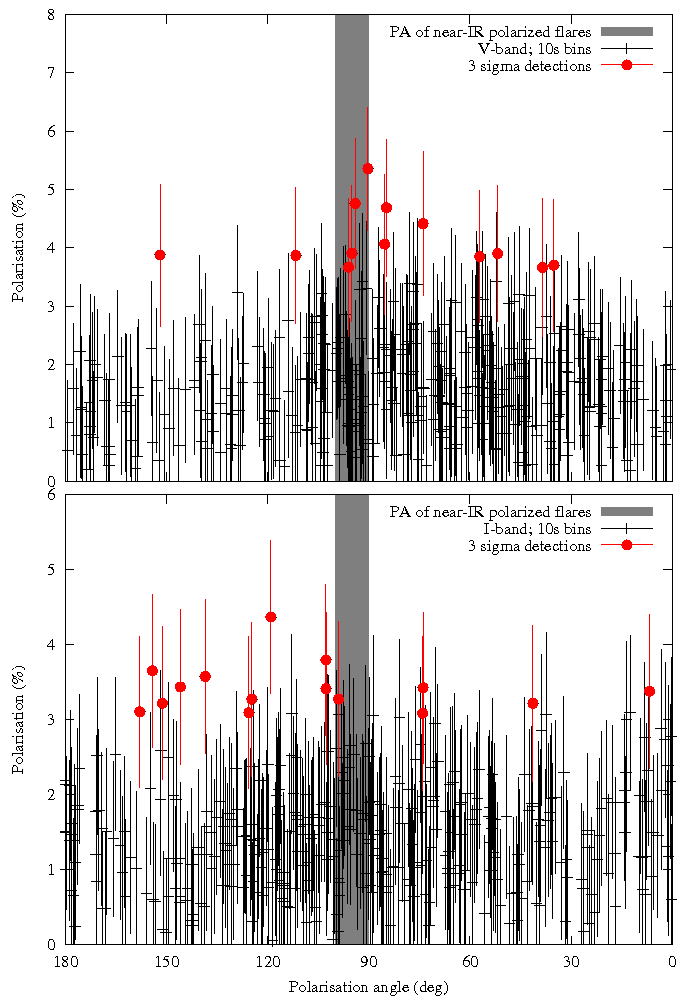}
\includegraphics[width=7.5cm,angle=0]{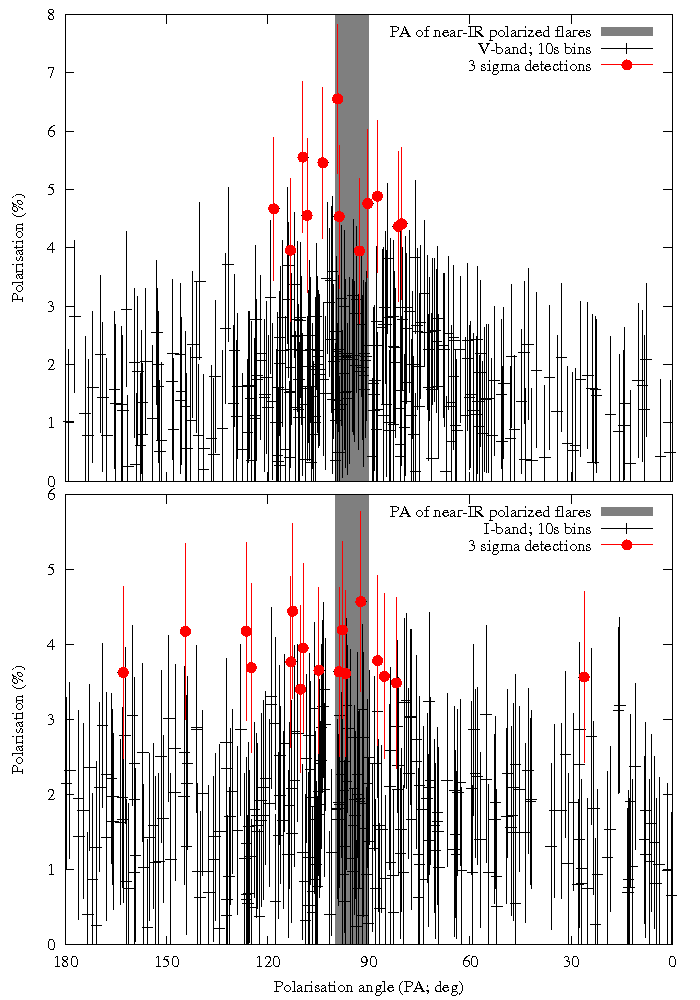}
\caption{The polarization PA of the significant (3$\sigma$; red circles) and insignificant ($< 3 \sigma$; black crosses) detections of LP, on Feb 19$^{th}$ (left) and Feb 22$^{nd}$ (right). V-band data are in the upper panels and I-band in the lower panels.}
\end{figure}

In Fig. 5 the X-ray hardness--intensity diagrams are plotted. We compared the data on Feb 19$^{th}$ (black data in left panel) and Feb 22$^{nd}$ (black data in right panel) with those of the period Feb 19--25$^{th}$, and compared the X-ray power spectra with those in the literature \cite{vandet97}. We find the source was in the `flaring branch' on 19$^{th}$ and the `normal branch' on Feb 22$^{nd}$ of the NS Z-source classifications.

To test whether these optical polarization flares could be real, we plot the PA of the significant (at the 3$\sigma$ level) detections of LP (in red) and the PA of the non-detections (black data) in Fig. 6 (using 10-sec binned data). All non-detections are $\sim$ evenly distributed in PA. On Feb 19$^{th}$ (left panels), almost all 3$\sigma$ detections of LP appear to have random PAs (again approximately evenly distributed) but on Feb 22$^{nd}$ (right panels) most of the 3$\sigma$ detections are clustered around PA $\sim$ 100$^{\circ}$. When we detected polarized NIR flares from the jet in 2006 (section 2), their PA was also $\sim$ 100$^{\circ}$ (shown as the grey regions in Fig. 6), which suggests the flares of LP could be real, and originate in the jets.

We see the polarized flares during the normal branch and not (significantly) during the flaring branch; this is consistent with the global picture of jet--disc coupling in NS Z-sources \cite{miglfe06}. The implication is that the optical emission from Sco X--1 is sometimes polarized by several per cent, but on timescales of 10 seconds or less. The jets appear to cause this transient polarization, and the flares could arise due to brief ordering of the predominantly tangled magnetic field near the jet base, or intermittency of the jets themselves.

\section{Conclusions}

The polarimetric signature of synchrotron-emitting jets can be detected in the infrared and optical in LMXBs. The results so far (from at least one BH source, GX 339--4 and one NS source, Sco X--1) suggest that a generally turbulent (only partially ordered), rapidly changing magnetic field is present near the base of the jets close to where they are launched. When significant polarization is measured, the position angle is consistent with a magnetic field aligned with (parallel to) the jet axis. The polarimetric variability timescales are at least as short as 10 seconds in Sco X--1, and the net polarization signal is diluted in exposures longer than this timescale. The short timescale of the flares of polarization places observational constraints on models for the jet launching process and the local magnetic field configuration. The magnetic field configuration, and hence polarization, cannot vary on timescales shorter than the light travel time across the emitting region. It is therefore very likely that a HTRA instrument with polarimetric capabilities on an extremely large telescope such as the planned European Extremely Large Telescope (E-ELT) will uncover the flux and polarization properties of these objects over \emph{all} plausible variability timescales.

\vspace{2mm}
\emph{Acknowledgements}
DMR thanks Steve Potter for help with the reduction and analysis of HIPPO data. Based on observations collected at the European Southern Observatory, Chile (under ESO Programme IDs 275.D-5062 and 281.D-5029), the United Kingdom Infrared Telescope (UKIRT; which is operated by the Joint Astronomy Centre on behalf of the Science and Technology Facilities Council of the U.K.) and the South African Astronomical Observatory (SAAO) 1.9-m Telescope. DMR acknowledges support from a Netherlands Organization for Scientific Research (NWO) Veni Fellowship.


\begin{thebibliography}{99}


\bibitem{bell10}Belloni T. M., 2010, in Belloni T., ed., Lect. Notes Phys., The Jet
Paradigm -- From Microquasars to Quasars, 794. Springer-Verlag, Berlin, p. 53

\bibitem{bjorbl82}Bj\"ornsson, C.-I., Blumenthal, G. R., 1982, ApJ, 259, 805

\bibitem{buxtve03}Buxton, M., Vennes, S., 2003, MNRAS, 342, 105

\bibitem{buxtba04}Buxton, M. M., Bailyn, C. D., 2004, ApJ, 615, 880

\bibitem{cardet89}Cardelli, J. A., Clayton, G. C., Mathis, J. S., 1989, ApJ, 345, 245

\bibitem{caseet10}Casella, P., et al., 2010, MNRAS, 404, L21

\bibitem{charet80}Charles, P. A., et al., 1980, ApJ, 237, 154

\bibitem{collet09}Collins, P. P., Shehan, B., Redfern, M., Shearer, A., 2009, in Polarimetry days in Rome: Crab status, theory and prospects, Proceedings of Science (arXiv:0905.0084)

\bibitem{corbfe02}Corbel, S., Fender, R. P., 2002, ApJ, 573, L35

\bibitem{coriet09}Coriat, M., Corbel, S., Buxton, M. M., Bailyn, C. D., Tomsick, J. A., K\"ording, E., Kalemci, E., 2009, MNRAS, 400, 123

\bibitem{dola84}Dolan, J. F., 1984, A\&A, 138, 1

\bibitem{dolata89}Dolan, J. F., Tapia, S., 1989, PASP, 101, 1135

\bibitem{dubuch06}Dubus, G., Chaty, S., 2006, A\&A, 458, 591

\bibitem{fabiet82}Fabian, A. C., Guilbert, P. W., Motch, C., Ricketts, M., Ilovaisky, S. A., Chevalier, C., 1982, A\&A, 111, L9

\bibitem{fendet09}Fender, R. P., Homan, J., Belloni, T. M., 2009, MNRAS, 396, 1370

\bibitem{gallet03}Gallo, E., Fender, R. P., Pooley, G. G., 2003, MNRAS, 344, 60

\bibitem{gandet10}Gandhi, P., 2010, MNRAS, in press (arXiv:1005.4685)

\bibitem{glioet98}Gliozzi, M., Bodo, G., Ghisellini, G., Scaltriti, F., Trussoni, E., 1998, A\&A, 337, L39

\bibitem{hasiva89}Hasinger, G., van der Klis, M., 1989, A\&A, 225, 79

\bibitem{hjelet90}Hjellming, R. M., et al., 1990, ApJ, 365, 681

\bibitem{homaet05}Homan, J., Buxton, M., Markoff, S., Bailyn, C. D., Nespoli, E., Belloni, T., 2005, ApJ, 624, 295

\bibitem{hyneet03}Hynes, R. I., et al., 2003, MNRAS, 345, 292

\bibitem{jonket04}Jonker, P. G., Nelemans, G., 2004, MNRAS, 354, 355

\bibitem{fomaet01}Fomalont, E. B., Geldzahler, B. J., Bradshaw, C. F., 2001, ApJ, 558, 283

\bibitem{gallet04}Gallo, E., Corbel, S., Fender, R. P., Maccarone, T. J., Tzioumis, A. K., 2004, MNRAS, 347, L52

\bibitem{kanbet03}Kanbach, G., Kellner, S., Schrey, F. Z., Steinle, H., Straubmeier, C., Spruit, H. C., 2003, SPIE, 4841, 82

\bibitem{miglfe06}Migliari S., Fender R. P., 2006, MNRAS, 366, 79

\bibitem{miglet06}Migliari, S., Tomsick, J. A., Maccarone, T. J., Gallo, E., Fender, R. P., Nelemans, G., Russell, D. M., 2006, ApJ, 643, L41

\bibitem{miglet07}Migliari, S., et al., 2007, ApJ, 670, 610

\bibitem{pottet08}Potter, S., et al., 2008, SPIE, 7014, 179

\bibitem{pottet10}Potter, S., et al., 2010, MNRAS, 402, 1161

\bibitem{russfe08}Russell, D. M., Fender, R. P., 2008, MNRAS, 387, 713

\bibitem{russet07}Russell, D. M., Fender, R. P., Jonker, P. G., 2007, MNRAS, 379, 1108

\bibitem{russet08}Russell, D. M., Altamirano, D., Lewis, F., Roche, P., Markwardt, C. B., Fender, R. P., 2008, The Astronomer's Telegram, 1586

\bibitem{russet10}Russell, D. M., Maitra, D., Dunn, R. J. H., Markoff, S., 2010, MNRAS, 405, 1759

\bibitem{rybili79}Rybicki, G. B., Lightman, A. P., 1979, Radiative Processes in Astrophysics. Wiley, New York

\bibitem{serk73}Serkowski, K., 1973, in Interstellar Dust and Related Topics, IAU Symp. 52, eds. J. M. Greenburg \& H. C. van de Hulst, Reidel, Dordrecht, p. 145

\bibitem{slowet09}S\l owikowska, A., Kanbach, G., Kramer, M., Stefanescu, A., 2009, MNRAS, 397, 103

\bibitem{schuet04}Schultz, J., Hakala, P., Huovelin, J., 2004, BaltA, 13, 581

\bibitem{shahet08}Shahbaz, T., Fender, R. P., Watson, C. A., O'Brien, K., 2008, ApJ, 672, 510

\bibitem{torret08}Torres, M. A. P., et al., 2008, ApJ, 672, 1079

\bibitem{vandet97}van der Klis, M., Wijnands, R. A. D., Horne, K., Chen, W., 1997, ApJ, 481, L97

\end{thebibliography}
\end{document}